\documentclass{article}
\usepackage{amsmath}
\usepackage[affil-it]{authblk}
\usepackage{ragged2e}
\usepackage{indentfirst}
\usepackage{xcolor}
\usepackage{hyperref}
\usepackage{PRIMEarxiv}
\usepackage[utf8]{inputenc} 
\usepackage[T1]{fontenc}   
\usepackage{hyperref}      
\usepackage{url}           
\usepackage{booktabs}      
\usepackage{amsfonts}      
\usepackage{nicefrac}     
\usepackage{microtype}     
\usepackage{lipsum}
\usepackage{fancyhdr}      
\usepackage{graphicx}      
\graphicspath{{media/}}     
\title{Decrypting the temperature field in flow boiling with latent diffusion models}
\author{UngJin Na\textsuperscript{a}, JunYoung Seo\textsuperscript{b}, Taeil Kim\textsuperscript{a}, ByongGuk Jeon\textsuperscript{c}, HangJin Jo\textsuperscript{a, b,}\thanks{Corresponding author, jhj04@postech.ac.kr}}
\begin{document}

\maketitle
\begin{center}
\textit{\small\textsuperscript{a}Department of Mechanical Engineering, Pohang University of Science and Technology (POSTECH),} \\
\textit{\small Pohang, Republic of Korea} \\
\textit{\small\textsuperscript{b}Division of Advanced Nuclear Engineering, Pohang University of Science and Technology (POSTECH),} \\
\textit{\small Pohang, Republic of Korea} \\
\textit{\small\textsuperscript{c}Korea Atomic Energy Research Institute, Daejeon, Republic of Korea} \\
\end{center}

\vskip 0.23in%

\begin{abstract}
This paper presents an innovative method using Latent Diffusion Models (LDMs) to generate temperature fields from phase indicator maps. By leveraging the BubbleML dataset from numerical simulations, the LDM translates phase field data into corresponding temperature distributions through a two-stage training process involving a vector-quantized variational autoencoder (VQVAE) and a denoising autoencoder. The resulting model effectively reconstructs complex temperature fields at interfaces. Spectral analysis indicates a high degree of agreement with ground truth data in the low to mid wavenumber ranges, even though some inconsistencies are observed at higher wavenumbers, suggesting areas for further enhancement. This machine learning approach significantly reduces the computational burden of traditional simulations and improves the precision of experimental calibration methods. Future work will focus on refining the model’s ability to represent small-scale turbulence and expanding its applicability to a broader range of boiling conditions.
\end{abstract}


\section{Introduction}
Flow boiling plays an important role in enhancing the performance of thermal management systems, including refrigeration, microelectronics cooling, nuclear power plants, and nuclear fission reactors \cite{carey2020phasechange, thome2004microchannels}. This phenomenon involves a fluid absorbing heat and undergoing a phase change from liquid to vapor, while supplied with the advection of the bulk flow, significantly boosting the heat transfer efficiency through the utilization of latent heat. The initiation of the phase change is known as the onset of nucleate boiling (ONB) \cite{collier1994convective}. However, when the liquid fails to rewet the surface, the surface becomes entirely covered by a vapor layer, leading to a significant reduction in heat transfer efficiency. This phenomenon is known as the departure from nucleate boiling (DNB) \cite{kandlikar2001chf}.

The heat transfer process between the ONB and the DNB points can be described using the RPI wall boiling model \cite{kurul1990multidimensional}. This model breaks down the total heat flux from the wall to the working fluid into three components: convective heat flux ($q_c$), quenching heat flux ($q_q$), and evaporation heat flux ($q_e$). The convective heat flux accounts for the heat transfer between the liquid phase and the heated wall surface not covered by bubbles. The quenching heat flux represents the cyclic averaged transient heat transfer as liquid refills the wall vicinity after bubble detachment. The evaporative heat flux is associated with the latent heat carried away by departing bubbles.

To estimate the heat transfer, visualization methods such as side-view shadowgraphs, total reflection, and infrared thermography have been utilized as major tools for collecting information on bubble dynamics \cite{amidu2018partitioning, jeon2022nucleate}. Additionally, Laser-Induced Fluorescence (LIF) techniques have been adopted to experimentally measure the spatial temperature of the fluids \cite{ghazvini2024fluorescence}. However, accurately estimating the actual temperature field remains extremely challenging. Furthermore, detailed computational models for subcooled boiling flow, such as the Eulerian-Eulerian framework, involve establishing mass, momentum, and energy conservation equations for each phase separately, which requires tremendous resources to calculate \cite{sato2017les}.

Recent advancements in machine learning, particularly in convolutional neural networks (CNNs), offer hope in addressing these challenges \cite{lecun2015deep}. CNNs have shown proficiency in capturing visual features, suggesting the potential to simplify complex experimental setups and data extraction processes \cite{krizhevsky2012imagenet}. Also, generative models, especially Generative Adversarial Networks (GANs), have shown promise in image processing tasks by learning the style of input data and generating corresponding images \cite{goodfellow2014gans}. However, GANs face challenges such as mode collapse and instability during training, limiting their effectiveness in modeling complex distributions \cite{radford2015dcgan}. Addressing these limitations, denoising diffusion probabilistic models (DDPMs) have emerged as a robust alternative for high-resolution image synthesis \cite{sohl2015unsupervised, dhariwal2021diffusion}. These models, built from a hierarchy of denoising autoencoders, have achieved impressive results in image synthesis without mode collapse and training instabilities, setting new benchmarks in conditional image generation and super-resolution tasks \cite{ho2020diffusion}. Latent Diffusion Models (LDMs) operate by first training an autoencoder to create a lower-dimensional representation of the data, reducing the dimensionality of the input to address computational inefficiencies, and then performing the training and sampling \cite{rombach2022ldm}.

This study presents a pioneering approach that utilizes LDMs for synthesizing temperature data in a flow boiling system. The model is trained to map phase field data to corresponding temperature field data. By integrating advanced conditioning mechanisms, the model learns and generates outputs even without explicitly known transfer functions between the input and output fields. The autoencoder produces understandable latents from ground truth information, which are later used for the denoising process. Our method provides a robust framework for generating high-fidelity thermal information of the field, which is critical for understanding complex thermal-hydraulic interactions during flow boiling processes. This capability makes LDMs a powerful tool for analyzing digitized bubble information and provides access to high-resolution information to a broader range of researchers and applications.

\section{Implementation}

\subsection{Dataset Utilization}
This study utilizes the numerical simulation results of flow boiling from the BubbleML dataset, acquired by the Flash-X software \cite{hassan2024bubbleml, dhruv2021gravity}. In particular, the phase field indicators and the temperature field data are utilized. To introduce this dataset, the simulation was conducted by solving the mass, momentum, and energy conservation equations for incompressible flow. A single-fluid approach is adopted, with variable properties based on the location of the interface. The momentum and energy equations are non-dimensionalized with reference quantities from the liquid phase defining the Reynolds number, Prandtl number, and Froude number. Temperature is scaled using $(T-T_{\text{wall}}) / (T_{\text{wall}}-T_{\text{bulk}})$. The liquid-vapor interface is tracked using a level-set advection equation. The fluid used is FC-72, known for its stability and insulation properties. The simulation has a domain resolution of $n_x=1344$, $n_y=160$, for the domain size of $l_x=29.4 \, \text{mm}$, $l_y=3.5 \, \text{mm}$. The non-dimensional constants for the fluid can be found in \cite{hassan2024bubbleml}. Simulation outputs are stored in HDF5 files with variables such as velocities, temperature, and signed distance function.

\subsection{Implementation of the LDMs}
In this study, we utilized LDMs for image-to-image translation. The key advantage of LDMs lies in their ability to separate the training process into two distinct phases: perceptual compression and semantic compression. In the first phase, the perceptual autoencoder eliminates high-frequency details while retaining essential perceptual features. In the second phase, the diffusion model learns the semantic composition of the data within this compressed latent space, derived from the autoencoder, significantly reducing computational complexity while maintaining high-fidelity image generation.

To elaborate, the operational procedure of LDMs initiates with the encoding of the original image $x_0$ into a latent representation $z_0$ using the encoder function $E$, such that $z_0 = E(x_0)$. Here, we used the vector-quantized variational autoencoder (VQVAE) for the effective compression of $x_0$ by extracting and encoding essential perceptual features into a lower-dimensional space, setting the stage for the subsequent diffusion processes \cite{vandoord2017vqvae}.

This encoded representation, $z_0$, is then subjected to a forward diffusion process where Gaussian noise $\epsilon \sim \mathcal{N}(0,I)$ is incrementally added over $T$ timesteps according to the equation: 
\[
z_t = \sqrt{\alpha_t} \, z_0 + \sqrt{1-\alpha_t} \, \epsilon.
\]
In this formula, $\alpha_t$ denotes the noise schedule that governs the variance of noise added at each timestep $t$.

Throughout the diffusion process, a denoising autoencoder is employed to systematically remove the noise, progressively transforming the noise into the latent representation. The reverse diffusion process aims to denoise $z_t$ back to $z_0$. During this stage, the model employs a neural network, parameterized by $\theta$, to predict the noise component $\epsilon_0$ at each timestep, represented as: $\hat{\epsilon} = \epsilon_0(z_t,t)$.

Finally, the denoised latent representation $z_0$ is passed through the decoder $D$ of the VQVAE to reconstruct the image, resulting in $\hat{x}_0 = D(z_0)$. This two-phase process efficiently utilizes the reduced dimensionality of the latent space to capture and synthesize the complex images.

\subsection{Neural Network Structure}
As the architecture of our LDM comprises two major parts: VQVAE and denoising autoencoders, the architecture of the hierarchical autoencoders is elaborated below.

\begin{figure}[hbt!]
\centering
\includegraphics[width=1\textwidth]{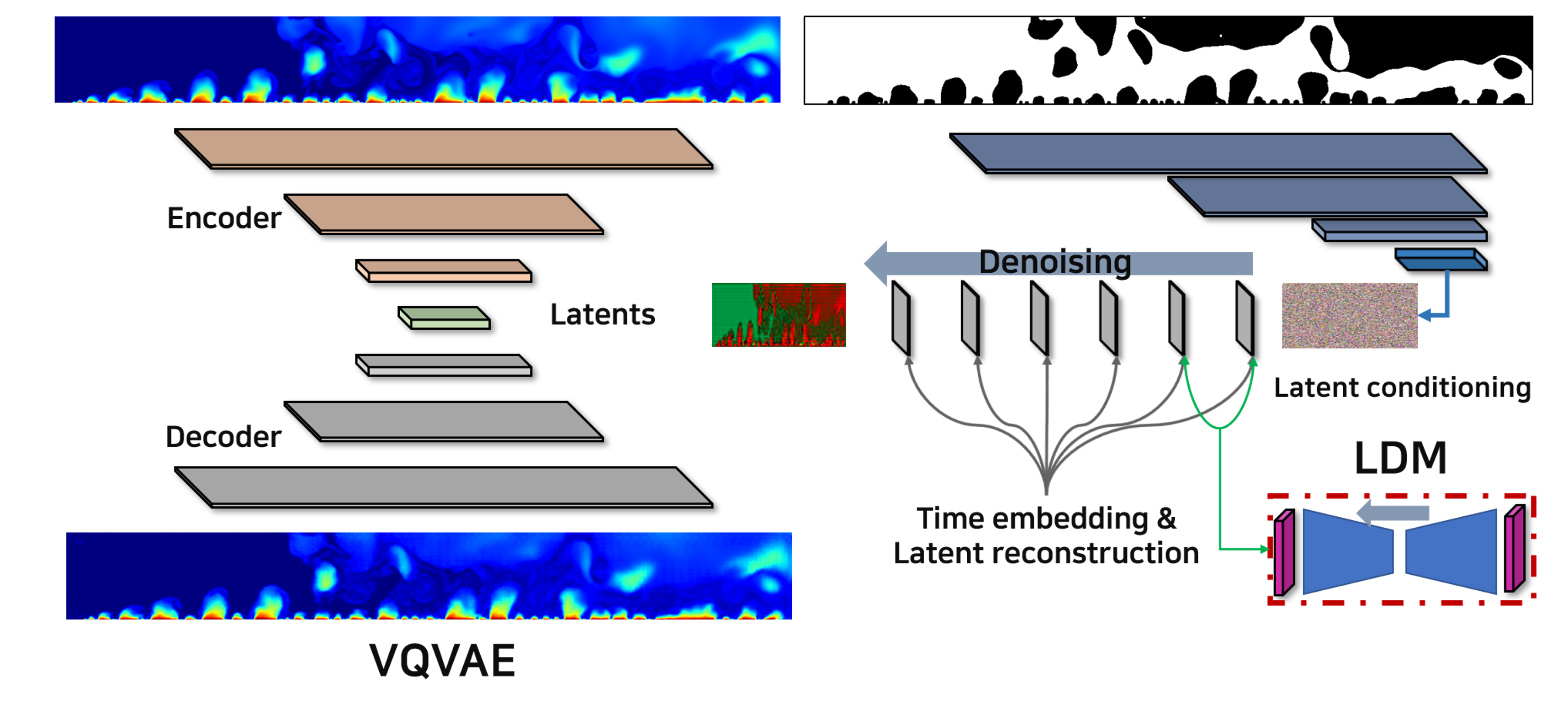} 
\caption{LDM Architeture}
\end{figure}

\subsubsection{VQVAE Architecture}
The input, sized at $160 \times 1344$, must be compressed into the latent size for efficient denoising. VQVAE extracts the latent representation, and upon recovery, the decoder layer generates the output image by decompressing the latent data. The architecture of VQVAE is U-Net-like, consisting of a downsampling encoding path, a bottleneck, and an upsampling decoding path. The encoding path transfers information to the decoding path through skip connections, where features are concatenated with corresponding upsampling layers to preserve and utilize low-level information. Each layer in the encoding path, bottleneck, and decoding path comprises a mix of convolution layers, group normalization, and SiLU activation functions.

In the encoding path, the encoder reduces the dimensionality of the source image. The number of feature channels increases while residual connections ensure the preservation of input information and efficient gradient propagation. Convolution operations determine the downsampling size of the input data. The specific design choices include the kernel size, stride, and padding as follows:

\begin{table}[h!]
\centering
\caption{Resizing convolutions of the VQVAE encoder}
\begin{tabular}{|c|c|c|c|}
\hline
Layer & Kernel Size & Stride & Padding \\
\hline
1 & (3, 4) & (1, 2) & 1 \\
\hline
2 & (4, 4) & (2, 2) & 1 \\
\hline
3 & (3, 4) & (1, 2) & 1 \\
\hline
4 & (4, 4) & (2, 2) & 1 \\
\hline
\end{tabular}
\end{table}

The bottleneck processes the compressed input data, applying additional convolutions to refine feature maps and incorporating residual connections to preserve information flow without changing the sample size.

The decoding path uses transposed convolutions for upsampling operations, increasing the spatial dimensions of the feature maps. By passing the reconstructed latent through the decoding path, the model transforms the denoised latent representation back into the image space, producing the final translated image. The specific design choices for this process include the kernel size, stride, and padding as follows:

\begin{table}[h!]
\centering
\caption{Resizing convolutions of the VQVAE decoder}
\begin{tabular}{|c|c|c|c|}
\hline
Layer & Kernel Size & Stride & Padding \\
\hline
1 & (4, 4) & (2, 2) & 1 \\
\hline
2 & (3, 4) & (1, 2) & 1 \\
\hline
3 & (4, 4) & (2, 2) & 1 \\
\hline
4 & (3, 4) & (1, 2) & 1 \\
\hline
\end{tabular}
\end{table}

\subsubsection{Denoising Autoencoder Architecture}
After the encoder of VQVAE, the noising-denoising processes are applied with the diffusion models. We have adopted a U-Net-like architecture for the diffusion model, which consists of an encoding path, a bottleneck, and a decoding path. The LDM receives an $84 \, (n_x) \times 40 \, (n_y)$ latent space, and the sample size does not change during the process.

The following network implementation is adopted for efficient noise prediction and removal. This supports various conditioning mechanisms, among which we chose the image conditioning of the latent reconstruction. During the conditional training, the conditional image is passed through a series of layers with the same structure as the encoder of the VQVAE, then concatenated along the channel dimension with the latent of the corresponding image. The concatenated image is then passed through each block layer.

For the denoising autoencoder, we have adopted the following array of convolutional layers so that the output information can be the same size as the given latent. During training, the timesteps are embedded using a sinusoidal embedding function and projected through linear layers. This embedding is added to the feature maps to incorporate temporal information, which is critical for tasks involving sequential data. This applies iterative denoising to learn the latent representation.

\begin{table}[h!]
\centering
\caption{The convolutions of the denoising autoencoder}
\begin{tabular}{|c|c|c|c|}
\hline
Layer & Kernel Size & Stride & Padding \\
\hline
1 & (3, 3) & (1, 1) & 1 \\
\hline
2 & (3, 3) & (1, 1) & 1 \\
\hline
3 & (3, 3) & (1, 1) & 1 \\
\hline
\end{tabular}
\end{table}

\subsection{Training Principles}
Before training, the dataset is prepared by locating HDF5 files, which are then grouped into conditioning files and output files. The dataset comprises 200 timesteps representing the experimental conditions, with images for the training and test datasets randomly shuffled at an 8:2 ratio. Data loaders are instantiated to manage the loading and preprocessing of images, and the models are transferred to GPU CUDA for processing. 
During the first training process, the VQVAE module is trained to learn $D(E(y_i)) = y_i$, where $y_i$ denotes the $i^{\text{th}}$ sequential temperature data. This process helps the autoencoder compress the input and reconstruct the output properly. Once the autoencoder training is complete, the denoising diffusion model receives latents $z_i = E(y_i)$ to repeat noise addition–denoising processes. As this study focuses on the conditioning of the latents, $z_{\text{cond}} = E_{\text{cond}}(x_i)$ is added to concatenate $z = [z_i \; z_{\text{cond}}]$. The training of the neural network models occurs while the architecture tries to denoise a random tensor of the same size as the given latent space, to reconstruct it.

The objective function of the denoising process in the DDPM is designed to optimize the details of the latent space, improving the quality of the generated images. This process is modeled as a Markov chain parameterized with variational inference, learning Gaussian transitions. By iteratively updating the model parameters through this optimization process, the DDPM effectively learns to generate realistic images that closely resemble the ground truth.

The training loop continues for 5000 steps for the VQVAE and 1000 steps for the LDM. The learning rate for VQVAE is $1 \times 10^{-5}$ and $5 \times 10^{-6}$ for LDM. For the diffusion process, the number of timesteps is set to 1000, with the starting $\beta$ at $8.5 \times 10^{-4}$ and the ending value at $1.2 \times 10^{-2}$. In each epoch, the noisy images and conditioning inputs are fed into the U-Net model to predict the noise. A linear noise scheduler is established during the forward diffusion process to manage the addition of random noise based on the current timestep. Then, the mean squared error (MSE) loss between the predicted noise and the actual noise is computed, gradients are calculated, and the optimizer updates the model's parameters. The VQVAE model remains frozen during the training of the LDM to ensure its parameters are not updated. The LDMs are trained through the integration of phase information and temperature images, which align with the actual values of temperature. The entire model implementation and training are conducted using Python 3.10.12, PyTorch 2.0.0, CUDA 11.7.64, cuDNN 8.9.7, and an RTX 4090 Ti on an Ubuntu 22.04 system.

\section{Results}

\subsection{Training of VQVAE}
The reconstruction of the temperature field has been achieved through the process of latent extraction using the VQVAE. As shown in the latent images, this method effectively captures the bubble structures within the field. Additionally, from the reconstructed temperature field, the VQVAE generates the overall temperature distribution with fine details. This demonstrates that the autoencoder can generate the temperature field accurately when provided with the proper latent representations, which will later be obtained through the denoising of latent diffusion. This process ensures that the intricate patterns and structures within the temperature field are preserved and accurately represented.

\begin{figure}[hbt!]
\centering
\includegraphics[width=1\textwidth]{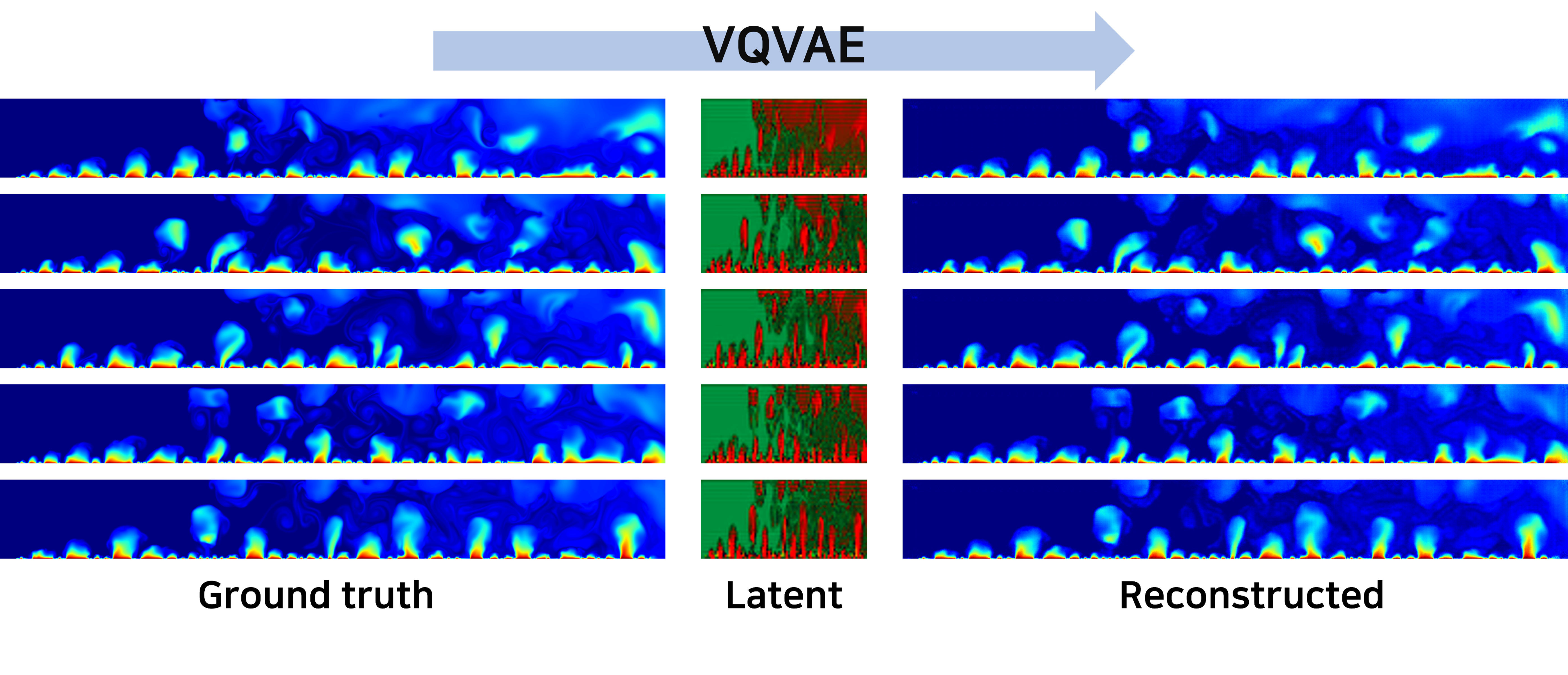} 
\caption{Reconstruction of ground truth via latent extraction}
\end{figure}

\subsection{Snapshots of the Generated Temperature Field}
The temperature field was generated using the LDM from the phase indicator map, where black represents the bubbles and white represents the liquid. This binary phase map serves as the input for the LDM to generate the latent space conditioning.

\begin{figure}[hbt!]
\centering
\includegraphics[width=1\textwidth]{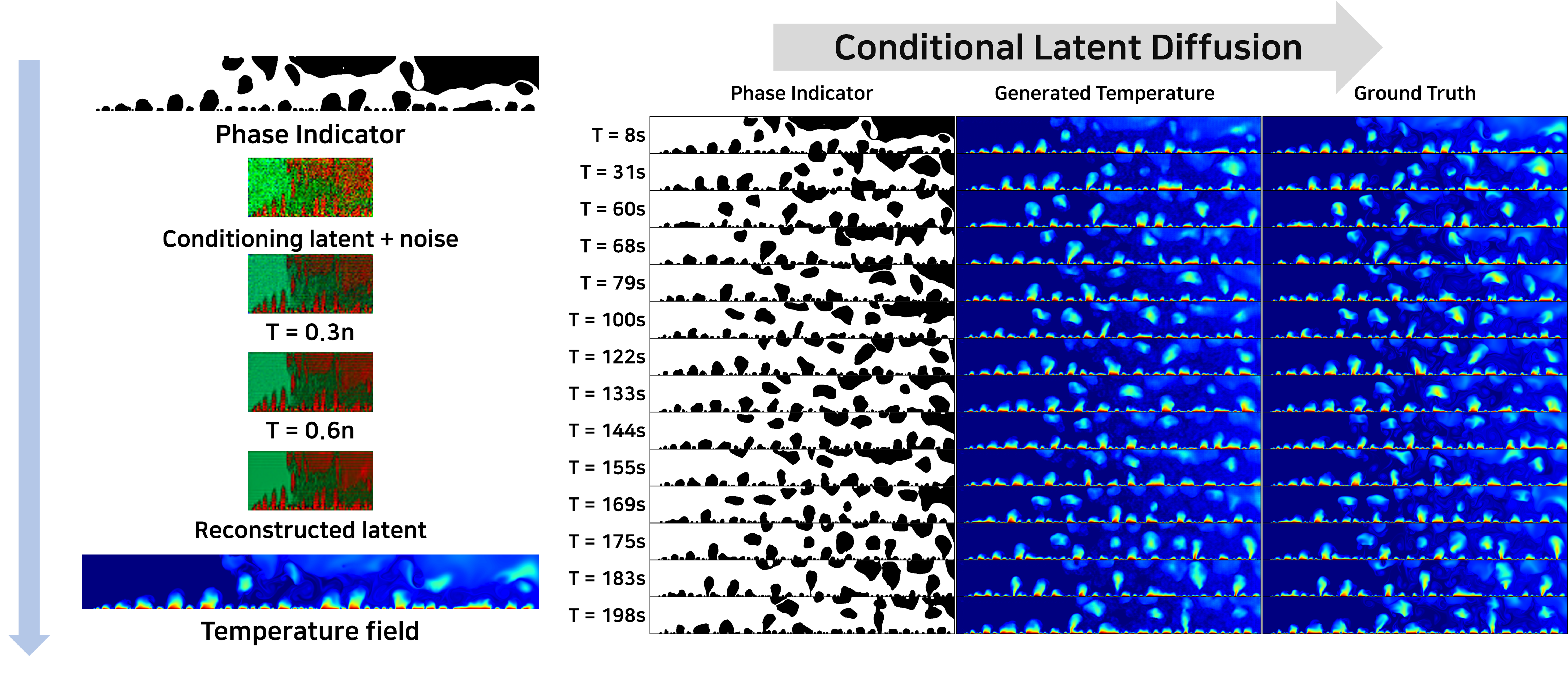} 
\caption{Temperature generation from phase indicator via LDM}
\end{figure}

As shown in Fig. 3, the model can generate a highly detailed temperature field. This detail is particularly evident within the bubbles and at the bubble interfaces, demonstrating a clear correlation between the phase boundaries and the temperature variations. Also, the passive scalar energy spectrum was analyzed, which follows a scaling relationship given by:
\[
E_T(k) \propto k^{-5/3}
\]
where \( E_T(k) \) represents the energy spectrum as a function of the wavenumber \( k \). This scaling highlights the distribution of energy across different scales in the turbulent regime.

The Fourier transform of the temperature field \( T(x) \), after subtracting the spatial mean temperature \( \langle T(x) \rangle \), is expressed as:
\[
\hat{T}(k) = \mathcal{F}\{T(x) - \langle T(x) \rangle\}
\]
where \( \mathcal{F} \) denotes the Fourier transform operator. This transformation converts the spatial temperature variations into their corresponding wavenumber components, facilitating spectral analysis.

To compute the power spectral density (PSD), the following equation is used:
\[
E_T(k) = \frac{1}{N_k} \sum_{|k| \in [k \pm \Delta k / 2]} |\hat{T}(k)|^2
\]
Here, \( N_k \) represents the number of wavenumbers within the interval \( [k \pm \Delta k / 2] \), and \( |\hat{T}(k)|^2 \) corresponds to the magnitude of the Fourier-transformed temperature field.
\begin{figure}[hbt!]
\centering
\includegraphics[width=1\textwidth]{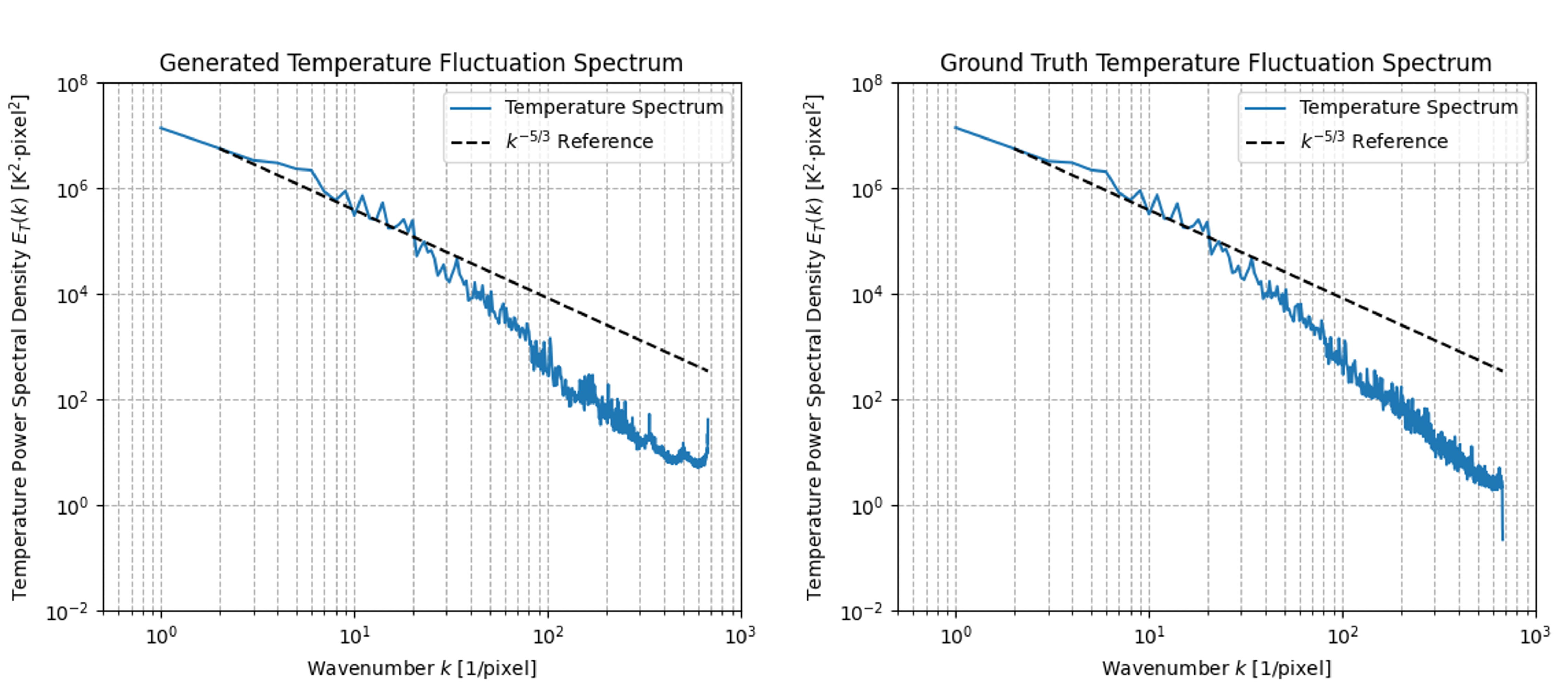} 
\caption{Temperature Fluctuation Spectrum}
\end{figure}

Over time, the spectral density of the generated data showed good agreement with the ground truth data in the low-to-mid wavenumber range. However, deviations were observed in the high-wavenumber range, indicating that the model struggles to capture small-scale turbulence accurately. This highlights the need for further refinement of the model to better represent the dynamics of small-scale turbulent structures. Moreover, as the temperature field is a strong indicator of bubble formation, growth, and departure, accurate prediction of thermal interactions between bubbles and the surrounding liquid, as well as the overall heat transfer efficiency, allows for improvements in pre-existing mechanistic models. 

\section{Discussion and Conclusion}
This study explored the potential of advanced machine learning techniques to enhance our understanding of flow boiling heat transfer while highlighting the effectiveness of LDMs in simplifying the training process and facilitating rapid image generation. This integrated approach allows for detailed visualization and analysis of the boiling process, providing insights that are difficult to achieve with traditional experimental methods alone.

We have demonstrated that latent diffusion models can be trained to convert segmented interface images of two-phase indicators into corresponding temperature distributions. Our findings suggest that perceptual loss alone can serve as an effective reconstruction mechanism, provided that high-fidelity data is available and the model is capable of processing it efficiently. Furthermore, this highlights the potential benefits of digitizing the two-phase domain using various tools, such as Mask R-CNN, to enhance analysis and reconstruction accuracy. This involves identifying and segmenting the interface in flow boiling systems. From generic shadowgraphs, by obtaining the sign function of these interfaces, we can create a digital representation of the spatial boundaries within the boiling system. Combining this data with image translation techniques opens new avenues for accurately mapping the temperature fields.

The implications of this approach are significant. Firstly, it offers a method to bypass the computational complexity of direct numerical simulations typically required for modeling subcooled boiling flow. By leveraging the strengths of machine learning, we can achieve high-fidelity results with reduced computational resources. Secondly, this technique provides a basis for thermal field estimations that are congruent with visualization data, which is crucial for optimizing thermal management systems in applications where only visual probes can be applied. The dataset generated from this approach can serve as a valuable calibration database for experimental techniques such as LIF. Such experimental techniques, used to measure spatial temperature distributions, can benefit greatly from a well-calibrated dataset to improve their reliability. By providing detailed field information on temperature, our machine learning models can enhance the calibration process, ensuring that experimental measurements are more precise and aligned with actual conditions.

Future research should focus on refining these machine learning models and expanding their application to various boiling conditions and configurations. Additionally, integrating real-time data acquisition systems with these predictive models could lead to the development of advanced control systems for thermal management, capable of dynamically adjusting operational parameters to optimize performance.

\bibliographystyle{unsrt}  
\bibliography{reference}

\end{document}